\documentclass[aps,prd,onecolumn,nofootinbib,superscriptaddress]{revtex4}
\usepackage{graphicx}
\pdfoutput=1
\usepackage{epsfig}
\usepackage{amsmath}
\usepackage{amsfonts}
\usepackage{amssymb}
\usepackage{color}
\usepackage{dcolumn}
\usepackage{hyperref}
\usepackage{multirow}
\usepackage{booktabs}
\usepackage{MnSymbol,wasysym}
\usepackage{braket,diagbox}
\usepackage{eurosym}
\usepackage{calrsfs}
\usepackage{subfigure}

\providecommand{\U}[1]{\protect\rule{.1in}{.1in}}
\newcommand{\red}[1]{\textcolor[rgb]{1.00,0.00,0.00}{#1}}

\newcommand{\f}{\begin{equation}}
\newcommand{\ff}{\end{equation}}
\newcommand{\fa}{\begin{eqnarray}}
\newcommand{\ffa}{\end{eqnarray}}

\begin{document}
\baselineskip=0.5cm
\title{Using precession and quasiperiodic oscillations to constrain a rotating regular black hole}

\author{Meng-He Wu}
\email{mhwu@njtc.edu.cn} 
\affiliation{College of Physics and Electronic Information Engineering, Neijiang Normal University, Neijiang 641112, China}

\author{Hong Guo}
\email{guohong@ibs.re.kr}
\affiliation{Particle Theory and Cosmology Group, Center for Theoretical Physics of the Universe,
Institute for Basic Science (IBS), Yuseong-gu, Daejeon, 34126, Republic of Korea}

\author{Xiao-Mei Kuang}
\email{xmeikuang@yzu.edu.cn}
\affiliation{ Center for Gravitation and Cosmology, College of Physical Science and Technology, Yangzhou University, Yangzhou 225002, China}

\begin{abstract}
\baselineskip=0.5cm
In this paper, we investigate the frame-dragging effect on an accretion disk and test gyroscope orbiting around a rotating regular black hole with a Minkowski core. Firstly, we perturb a bound timelike circular orbit around the black hole, and analyze the periastron precession and  Lense-Thirring (LT) precession frequencies of the orbit's epicyclic oscillations. Since these epicyclic oscillations can be used to explain the quasiperiodic oscillations (QPOs) phenomena of the accretion disc around this rotating regular black hole, we then employ the Markov Chain Monte Carlo (MCMC)  simulation to fit our theoretical results with five QPOs events (GRO J1655-40,  GRS 1915+105, XTE J1859+226, H1743-322 and  XTE J1550-564). The simulations give the relevant physical parameter space of the black hole, including  the characteristic radius $r$, the mass related parameter $M$, the spinning parameter $a$ and the quantum gravity effect $\alpha$. The results give
the constraint on the quantum effect parameter, with an upper limit  $\alpha/M^{2/3} < 0.60$ at the $95\%$ C.L., which is tighter than $<0.7014$ in our pervious study within static case.
Then, we theoretically explore the LT precession frequency, geodetic precession frequency, and the general spin precession frequency of a test gyro attached to a stationary observer in this black hole background. We find that the quantum gravity effect suppresses the precession frequencies comparing against those  in Kerr black hole,  further providing a theoretical diagnostic of the potential quantum gravity effect.


{\bf PACS number(s):} 04.70.-s,04.70.Dy,04.80.Cc
\end{abstract}

\maketitle

\newpage
\tableofcontents

\section{Introduction}
Singularities are inevitable predictions of general relativity (GR), signaling the breakdown of the classical theory in the ultraviolet regime. It is generally believed that a complete theory of quantum gravity will ultimately resolve these singularities. Long before such a theory was established, however, Bardeen proposed the first explicit model of a static regular black hole \cite{Bardeen:1968}, which features a non-singular core rather than a central curvature singularity. Later, it was shown that this solution can be derived from Einstein gravity coupled to a nonlinear magnetic monopole source \cite{Ayon-Beato:2000mjt}, thereby providing a physical basis for the idea of singularity-free black holes. Since then, many other regular black hole models have been constructed.
Existing constructions of regular black holes can be broadly divided into two categories. The first is semiclassical in nature, obtained by solving Einstein or generalized gravity equations with exotic or effective matter sources, such as nonlinear electrodynamics or noncommutative geometry inspired distributions \cite{dymnikova1992vacuum, Nicolini:2005vd,Balakin:2016mnn,Roupas:2022gee}. The second is quantum-motivated, in which regular black holes are interpreted as effective geometries incorporating quantum corrections to classical singular solutions, as suggested by loop quantum gravity, asymptotic safety, and related approaches \cite{Borde:1996df, Bonanno:2000ep, Gambini:2008dy, Perez:2017cmj, Brahma:2020eos,Gan:2022oiy}.

Regular black holes thus provide valuable theoretical laboratories. They allow us to explore semiclassical and quantum aspects of black holes, and examine potential observational signatures that could reveal deviations from classical singular geometries. In this sense, regular black holes serve as effective bridges between GR and candidate quantum gravity theories. For comprehensive reviews, reader can refer to \cite{Torres:2022twv,Lan:2023cvz}.
Therefore, it is interesting to study the phenomena to distinguish the regular black holes and normal black holes, especially in the strong gravity sector, the properties of which can be encoded  in the observable features or effects of the surrounding sectors around the central objects.

On one hand, the spectrum emitted by accretion flows around compact objects provides a crucial observational window into the strong-gravity regime. Since considerable radiation originates in the deep gravitational potential well of these systems, the study of such emission offers a unique opportunity to probe the near-horizon structure of black holes. Although no radiation can emerge directly from within the event horizon, valuable information can be extracted from the surrounding accretion flow, particularly from the accretion disks \cite{Bardeen:1972fi}. These disks typically produce a soft X-ray continuum, whose spectral features, such as the frequency range and thermal cutoff, can be used to infer the innermost disk radius. In standard GR, this radius coincides with the innermost stable circular orbit (ISCO), which encodes fundamental information about the nature of the central object.
The idea of using X-ray variability to test the inner accretion region dates back to the early 1970s, when it was proposed in \cite{Syunyaev1972} that rapid modulations in the flux could reflect geodesic motion in the strong-field regime. This idea gained renewed relevance with the discovery of quasi-periodic oscillations (QPOs) in X-ray binaries. These are characteristic peaks in the Fourier power spectrum of noisy X-ray light curves from accreting compact objects \cite{Lewinbook, Motta:2016vwf,eXTP:2018kpm}. QPOs are commonly classified into high-frequency (HF, $0.1$–$1,\text{kHz}$) and low-frequency (LF, $<0.1,\text{kHz}$) categories \cite{Stella:1997tc, Stella:1999sj}. Remarkably, HF QPOs often appear in pairs (upper and lower frequencies), with ratios clustering near $3:2$ in black hole microquasars \cite{Kluzniak:2001ar}. The upper frequencies are typically close to the orbital frequencies of test particles near the ISCO, reinforcing the idea that QPOs encode direct signatures of strong-field orbital dynamics.

A variety of theoretical models have been developed to explain QPOs, ranging from purely geodesic interpretations to resonant and nonlinear oscillation frameworks. The first suggestion that QPOs might serve as probes of strong gravity was based on the geodesic motion of test particles \cite{Kluzniak1990}. Since then, epicyclic motion, characterized by orbital, radial, and vertical oscillation frequencies, has been extensively investigated as a diagnostic tool for modeling HF QPOs. Matching the observed frequency patterns with theoretical predictions allows one to constrain both the parameters of the compact object and possible deviations from GR \cite{Zhang:2009gn,Bambi:2012pa,Bambi:2013fea,Maselli:2014fca,Jusufi:2020odz,Ghasemi-Nodehi:2020oiz,
Chen:2021jgj,Allahyari:2021bsq,Deligianni:2021ecz,Deligianni:2021hwt,Jiang:2021ajk,Banerjee:2022chn,
Liu:2023vfh,Riaz:2023yde,Rayimbaev:2023bjs,Abdulkhamidov:2024lvp,Jumaniyozov:2024eah,Guo:2025zca,Wu:2025ccc} and references therein. Moreover, more precise and higher resolution X-ray observations are expected to significantly sharpen these constraints. In particular, the Chinese Insight-HXMT mission (Hard X-ray Modulation Telescope) \cite{Lu:2019rru,Guo:2020ulb} has already improved the quality of X-ray timing measurements, while the next generation Einstein Probe mission \cite{yuan2018einstein,Yuan:2025cbh}, dedicated to time-domain X-ray astronomy, promises to provide stringent new tests of accretion physics and the nature of central compact objects in strong gravity.

On the other hand, for timelike orbits around a rotating compact object, two remarkable relativistic effects are predicted in GR,  the geodetic (or de Sitter) precession \cite{deSitter:1916zz} and the Lense-Thirring (LT) precession \cite{Lense:1918LT}. The geodetic effect arises purely from spacetime curvature generated by the central mass, whereas the LT effect is a consequence of frame dragging due to the rotation of the central body. A natural way to probe these effects is to study the behavior of a test gyroscope moving around a central object, since a spinning gyro tends to maintain its axis aligned with respect to reference stars, while spacetime curvature and frame dragging induce precessions of its spin direction. Both effects have been measured in Earth's gravitational field by several satellite experiments, such as the Gravity Probe B (GPB) where a satellite carrying four gyroscopes and a telescope orbits 650 km above the Earth~\cite{Everitt:2011hp}, the Laser Relativity Satellite (LARES) which was launched to measure the frame dragging effect with an accuracy of about $10^{-2}$~\cite{Capozziello:2014mea}, the LAGEOS satellites where the satellite acts as the particle moving around the earth \cite{Ciufolini:2004rq}, the Earth-based Gyroscopes IN General Relativity (GINGER) experiment which can measure Geodesic and Lense-Thirring effects with an uncertainty of 1 part in $10^4$ and $10^3$~\cite{DiVirgilio:2021ziy}, and future measurements of relativistic frame dragging and geodetic precessions by Gravity Probe
Spin (GPS) Satellite which will include the intrinsic spin of the electron~\cite{Fadeev:2020gjk}.
All those provide one of the most direct tests of frame dragging in nature.

From the theoretical aspect in strong field regime, geodetic precession in Schwarzschild and Kerr spacetimes was studied in detail in \cite{sakina1979parallel}. The LT effect is inherently more intricate. In the weak-field approximation, the LT precession frequency scales linearly with the spin parameter of the central object and falls off as $r^{-3}$, with $r$ the orbital distance from the rotating source \cite{2009Hartle}. In the strong gravity regime, however, the precession frequency exhibits much richer behavior and depends sensitively on the nature of the central object. Studies have examined LT precession in Kerr black holes \cite{Chakraborty:2013naa,Bini:2016iym}, their generalizations \cite{Chakraborty:2012wv,Iyer:2025ccd,Wang:2025fuw}, as well as in more exotic rotating spacetimes\cite{Jia:2023see} such as traversable wormholes \cite{Chakraborty:2016oja} and rotating neutron stars \cite{Chakraborty:2014qba}. These investigations show that frame dragging provides a potential diagnostic of the underlying spacetime geometry.
Moreover, it has been proposed in \cite{Chakraborty:2016mhx} that the spin precession of a test gyroscope could distinguish a Kerr naked singularity from a Kerr black hole, since their LT frequencies behave differently in the strong-field region. This idea has been further explored in various extensions of GR, including modified gravity scenarios \cite{Rizwan:2018lht,Rizwan:2018rgs,Pradhan:2020nno,Solanki:2021mkt,Wu:2023wld,QiQi:2024dwc,Zhen:2025nah,Zahra:2025fvq}, reinforcing the view that gyroscopic precession can serve as a powerful tool to probe the fundamental nature of compact rotating objects.

Thus, one aim of this paper is to investigate the epicyclic motion of a test particle and its application to the observed data of QPOs in the vicinity of the rotating counterpart of the regular black hole with Minkowski core \cite{Ling:2021olm}.  We shall  use of Markov Chain Monte Carlo (MCMC) algorithm \cite{Foreman-Mackey:2012any} to fit the theoretical predictions for the QPO frequencies to the observational X-ray data of  GRO J1655-40,  GRS 1915+105, XTE J1859+226, H1743-322 and  XTE J1550-564 \cite{Strohmayer:2001yn}, respectively, which will provide parameter constraints  of the rotating regular black hole with Minkowski core. The other aim is to examine the spin precession frequency of a test gyroscope associated with a stationary observer in the rotating regular black hole spacetime. We shall analyze how the quantum effect affect the  de Sitter precession and the LT  precession of the test gyro, and evaluate there deviations from Kerr black hole in GR.

The remaining of this paper is organized as follows. In section \ref{sec:motion}, We briefly review a rotating regular black hole with a Minkowski core, and then construct the timelike geodesic motion in the background. In section \ref{sec:particle}, we firstly study the dynamics of epicyclic motion of timelike particles and explore how the quantum effect influence the epicyclic frequencies. Then, we use the MCMC simulations to determine the best-fit value of the black hole parameters with the use of observational data from five QPO event. In section \ref{sec:gyro}, we consider the geodesic motion of a test gyro attached to a stationary observer in this rotating regular black hole, and theoretically study the effect of quantum correction on the gyro's LT precession frequency, geodetic precession frequency, and the general spin precession frequency. The final section contributes to our conclusions and discussions.

\section{Timelike geodesic motion in a regular black hole spacetime}\label{sec:motion}

Now, let us briefly introduce the regular black hole which is characterized by the sub-Planckian curvature with a Minkowski core.
The spherically symmetric metric has the form as~\cite{Ling:2021olm}
\begin{equation}\label{metric:s}
\begin{aligned}
ds^2 = -f(r)\, dt^2 + f(r)^{-1} dr^2 + r^2 \left(d\theta^2 + \sin^2\theta\, d\phi^2 \right),
\end{aligned}
\end{equation}
where the metric function reads
\begin{equation}
\begin{aligned}\label{Eq:fr}
f(r) \equiv 1 - \frac{2m(r)}{r}=1-\frac{2M}{r}e^{-\alpha^n M^\gamma / r^n}.
\end{aligned}
\end{equation}
Here $M$ denotes the mass of the regular black hole, while $\alpha$ describes the deviation from the Newtonian potential and play the role of the quantum gravity effect.
Both $\gamma$ and $n$ are dimensionless parameters that must satisfy the condition $n/3\leq \gamma \leq n$ with $n\geq 2$ in order to ensure the formation of the event horizon as well as the existence of the Kretschmann scalar curvature at sub-Planckian levels~\cite{LingLingYi:2021rfn,Zeng:2022yrm,Tang:2025mkk}.
In this paper, we consider $\gamma=1$ and $n=3$ for convenience, which correspond to an asymptotical Hayward black hole at large scales.
Since this regular black hole presents a quantum gravity effect in classic black hole metric, so it  has recently attracted considerable attention to investigate the quantum gravity modifications to black hole phenomena, such as quasinormal modes~\cite{Tang:2024txx,Zhang:2024nny}, Hawking radiation~\cite{Dong:2025duz}, as well as the potential astrophysical signal studies like black hole shadows~\cite{Meng:2023uws,Xiong:2025hjn}, accretion disk~\cite{Zeng:2023fqy} and gravitational lensing \cite{Sodejana:2024mus,Guo:2024svn}.
Those studies addressed that the regular black holes indeed provide a platform to use the astrophysical observations to test potential quantum modifications of gravity.

However, it is commonly accepted that  the astronomical black holes are generally more consistent
with the rotating (Kerr) solution within the classical framework.
Therefore, adopting the rotating  model
will facilitate more straightforward comparisons with
observational results. Thus, we construct the stationary and axisymmetric counterpart of the regular black hole \eqref{metric:s} by modified Newman-Janis algorithm \cite{Newman:1965tw,Azreg-Ainou:2014nra,Azreg-Ainou:2014pra}, and in Boyer-Lindquist coordinates, the rotating regular black hole has the metric \cite{Ling:2022vrv}
\begin{equation} \label{eq:matric_r}
\begin{aligned}
ds^2 =& g_{tt} dt^2 + g_{rr} dr^2 + g_{\theta\theta} d\theta^2 + g_{\phi\phi} d\phi^2 + 2g_{t\phi} dt\, d\phi \\
=& -\left(1 - \frac{2m(r)r}{\Sigma} \right) dt^2
- \frac{4a m(r) r \sin^2 \theta}{\Sigma} dt\, d\phi
+ \frac{\Sigma}{\Delta} dr^2 \\
& + \Sigma\, d\theta^2
+ \left(r^2 + a^2 + \frac{2a^2 m(r) r \sin^2 \theta}{\Sigma} \right) \sin^2 \theta\, d\phi^2,
\end{aligned}
\end{equation}
where
\begin{equation}
\begin{aligned}
\Delta = r^2 - 2m(r)r + a^2, \quad \Sigma = r^2 + a^2 \cos^2 \theta.
\end{aligned}
\end{equation}
This metric describes a Kerr-deformed regular black hole spacetime associated with the quantum gravity effect.
When $\alpha$ vanishes, the metric reduces to the standard Kerr solution in GR.
The parameter $a$ represents the black hole spin, and its vanishing results in the metric returning to the static limit~\eqref{metric:s}.

In the following discussion, we focus on the inertial frame-dragging effects by analyzing the epicyclic oscillations of particles on the accretion disk as well as the precession of the gyroscopes in strong field regime of this stationary regular black hole spacetime.
To proceed, we consider a test particle and a test gyroscope whose precession frequency can be determined from the timelike geodesic equations.
Since this spacetime admits two killing vectors, namely $\partial_t$ and $\partial_\phi$, the conserved energy $\mathcal{E}$ and the axial component of the angular momentum $L_z$ can be expressed as
\begin{equation} \label{eq:5}
\mathcal{E} = -g_{tt} \dot{t} - g_{t\phi} \dot{\phi}, \qquad
L_z = g_{t\phi} \dot{t} + g_{\phi\phi} \dot{\phi},
\end{equation}
where the dot represent the derivatives with respect to the affine parameter $\lambda$.
The Hamilton-Jacobi equation is introduced for a particle with rest mass $\mu$
\begin{equation}\label{eq:H}
\mathcal{H} = -\frac{\partial S}{\partial \lambda} = \frac{1}{2} g_{\rho \nu  }
\frac{\partial S}{\partial x^\rho} \frac{\partial S}{\partial x^\nu} = \mu^2.
\end{equation}
Here $\mathcal{H}$ is the canonical Hamiltonian, and $x^{\mu}$ represent the coordinate components $(t,r,\theta,\phi)$ in this metric ansatz.
The Jacobi action $S$ can be expressed in a separable form
\begin{equation} \label{eq:S}
S = \tfrac{1}{2} \mu^2 \lambda - \mathcal{E} t + L_z \phi + S_r(r) + S_\theta(\theta).
\end{equation}
Substituting Eq.~\eqref{eq:matric_r} and Eq.~\eqref{eq:S} into Eq.~\eqref{eq:H}, one can obtain four first-order geodesic equations
\begin{equation}
\begin{aligned}
\Sigma \dot{t} &= a (L_z - a \mathcal{E} \sin^2 \theta)
+ \frac{r^2 + a^2}{\Delta} \left( (r^2 + a^2) \mathcal{E} - a L_z \right), \\
\Sigma \dot{\phi} &= \frac{L_z}{\sin^2 \theta} - a \mathcal{E}
+ \frac{a}{\Delta} \left( (r^2 + a^2) \mathcal{E} - a L_z \right), \\
\Sigma \dot{r} &= \Delta \left( \frac{d S_r}{dr} \right)
= \sqrt{ \left( (r^2 + a^2) \mathcal{E} - a L_z \right)^2
- \Delta \left( Q + (L_z - a \mathcal{E})^2 + \mu^2 r^2 \right) }, \\
\Sigma \dot{\theta} &= \frac{d S_\theta}{d\theta}
= \sqrt{ Q + \cos^2 \theta \left( a^2 (\mathcal{E}^2 - \mu^2)
- \frac{L_z^2}{\sin^2 \theta} \right) },
\end{aligned}
\end{equation}
with the Carter constant $\mathcal{C}$ defined by
\begin{equation}
\left( \frac{d S_\theta}{d\theta} \right)^2
+ \frac{(L_z - a \mathcal{E} \sin^2 \theta)^2}{\sin^2 \theta}
= -\Delta \left( \frac{d S_r}{dr} \right)^2
+ \frac{ \left( (r^2 + a^2) \mathcal{E} - a L_z \right)^2 }{ \Delta } = \mathcal{C}
\end{equation}
where $Q \equiv \mathcal{C} - (L_z - a\mathcal{E})^2$.
It can be seen that the geodesic equations in our regular black hole spacetime are similar to those in the Kerr case, with all effects of the regular black hole encoded in the metric function $\Delta$.
Subsequently, we will examine the orbital precession separately for the cases of a test particle and a test gyroscope.

\section{Precession of timelike bound orbits and quasi-periodic oscillations} \label{sec:particle}

In this section, we analyze the precessing orbits of a test particle and characterize both the LT precession and periastron precession near the rotating regular black holes using the three characteristic frequencies derived from orbital perturbations.
Furthermore, we incorporate QPOs events observed in X-ray binaries and employ MCMC methods to constrain the parameters of the rotating regular black holes.

\subsection{LT precession and periastron precession of test particles}
We first derive the the orbital, periastron precession, and nodal precession frequencies that describe QPOs of accretion around the rotating regular black hole in the context of the relativistic precession model (RPM).  In this model, the accretion disk is composed of matter moving along nearly circular orbits under the influence of the strong gravitational field of the central compact object, and QPOs observed in X-ray binaries are interpreted as a natural outcome of the motion of
matter around the compact object. The characteristic frequencies raised from small  perturbations away from circular motion are directly related to the motion of test particles orbiting in the accretion disk around this black hole. Its structure and emission properties are determined by the spacetime geometry, which encodes the gravitational potential and frame-dragging effects specific to the black hole metric. Thus, to investigate the fundamental oscillatory modes responsible for the observed QPOs, we shall analyze the equations of motion for a massive test particle in this background. This provides the theoretical basis for computing the corresponding orbital and precessional frequencies that are crucial for modeling QPO phenomena in relativistic accretion disks \cite{Kluzniak1990,Stella:1999sj}.
To guarantee the conservation of the test particle’s rest mass, the timelike geodesic is required to satisfy the normalization condition
\begin{equation}\label{eq:28}
g_{\mu \nu} \dot{x}^\mu \dot{x}^\nu = -1.
\end{equation}
Using the metric \eqref{eq:matric_r} and the expressions for $\dot{t}$ and $\dot{\phi}$ given by Eq.~\eqref{eq:5}, we can rewrite Eq.~\eqref{eq:28} into
\begin{equation}\label{eq:29}
g_{rr} \dot{r}^2 + g_{\theta\theta} \dot{\theta}^2 = V_\text{eff},
\end{equation}
where the effective potential $V_\text{eff}$ takes the form
\begin{equation}\label{eq:30}
V_\text{eff} = \frac{\mathcal{E}^2 g_{\phi\phi} + 2 \mathcal{E} L_z g_{t\phi} + L_z^2 g_{tt}}{g_{t\phi}^2 - g_{tt}g_{\phi\phi}} - 1.
\end{equation}
For convenience, we then focus our analysis on circular geodesic motion on the equatorial plane ($\theta = \pi/2$) at a fixed radius $r = r_0$, which imposes the following conditions on the effective potential
\begin{equation}\label{eq:31}
V_\text{eff}(r_0, \pi/2) = \mathcal{E},\quad \partial_r V_\text{eff}(r_0, \pi/2) = 0,\quad \partial_\theta V_\text{eff}(r_0, \pi/2) = 0.
\end{equation}
Subsequently, the conserved energy $\mathcal{E}$ and angular momentum $L_z$ of the particle in this orbit can be expressed as
\begin{equation}\label{eq:33}
\mathcal{E} = \left.\frac{g_{tt} + g_{t\phi} \Omega_\phi}{\sqrt{-g_{tt} - 2 g_{t\phi} \Omega_\phi - g_{\phi\phi} \Omega_\phi^2}}\right|_{r = r_0, \theta = \pi/2}, \quad
L_z = \left.\frac{g_{t\phi} + g_{\phi\phi} \Omega_\phi}{\sqrt{-g_{tt} - 2 g_{t\phi} \Omega_\phi - g_{\phi\phi} \Omega_\phi^2}}\right|_{r = r_0, \theta = \pi/2},
\end{equation}
where the corresponding particle's angular velocity, as measured by an observer at infinity, is given by
\begin{equation}\label{eq:32}
\Omega_\phi \equiv \frac{d\phi}{dt} =\left.\frac{ -g_{t\phi}\pm \sqrt{ g_{t\phi}^2 - g_{\phi\phi} g_{tt} } }{ g_{\phi\phi} }\right|_{r = r_0, \theta = \pi/2}=
-\left.\frac{\mathcal{E} g_{t\phi} + L_z g_{tt}}{\mathcal{E} g_{\phi\phi} + L_z g_{t\phi}}\right|_{r = r_0, \theta = \pi/2}.
\end{equation}
Noted that the symbols $\pm$ correspond to prograde and retrograde orbits, respectively. A prograde orbit refers to a trajectory in which the particle’s angular momentum is aligned with the spin of the black hole, whereas in a retrograde orbit, the angular momentum is anti-aligned. In this study, we restrict our attention to the prograde case.

If the test particle on a stable circular orbit is slightly perturbed, it undergoes oscillations in both the radial and vertical directions, characterized by the epicyclic frequencies $\Omega_r$ and $\Omega_\theta$, respectively.
These frequencies can be derived from the linearized equations of motion as~\cite{Ryan:1995wh,Doneva:2014uma}
\begin{equation}
\Omega_r = \left. \left[ \frac{1}{2g_{rr}} \left(
X^2 \partial_r^2 \left( \frac{g_{\phi\phi}}{g_{tt}g_{\phi\phi} - g_{t\phi}^2} \right)
- 2XY \partial_r^2 \left( \frac{g_{t\phi}}{g_{tt}g_{\phi\phi} - g_{t\phi}^2} \right)
+ Y^2 \partial_r^2 \left( \frac{g_{tt}}{g_{tt}g_{\phi\phi} - g_{t\phi}^2} \right)
\right) \right]^{1/2} \right|_{r=r_0,\,\theta=\pi/2},
\end{equation}
\begin{equation}
\Omega_\theta = \left. \left[ \frac{1}{2g_{\theta\theta}} \left(
X^2 \partial_\theta^2 \left( \frac{g_{\phi\phi}}{g_{tt}g_{\phi\phi} - g_{t\phi}^2} \right)
- 2XY \partial_\theta^2 \left( \frac{g_{t\phi}}{g_{tt}g_{\phi\phi} - g_{t\phi}^2} \right)
+ Y^2 \partial_\theta^2 \left( \frac{g_{tt}}{g_{tt}g_{\phi\phi} - g_{t\phi}^2} \right)
\right) \right]^{1/2} \right|_{r=r_0,\,\theta=\pi/2},
\end{equation}
where $X$ and $Y$ are defined by
\begin{equation}
X = g_{tt} + g_{t\phi} \Omega_\phi, \quad Y = g_{t\phi} + g_{\phi\phi} \Omega_\phi.
\end{equation}

\begin{figure}[h]
    \centering
    \includegraphics[width=0.7\linewidth]{FIG1.png}
    \caption{The radial profiles of $\nu_{\text{nod}}$ and $\nu_{\text{per}}$ for selected parameters. Panels (a) and (b) describe the behavior of $\nu_{\text{nod}}$ and $\nu_{\text{per}}$, respectively, for $a=0$ (solid), $0.3M$ (dashed), $0.5M$ (dot dashed) with fixed $\alpha=0.6M^{2/3}$. Conversely, panels  (c) and  (d) depict $\nu_{\text{nod}}$ and $\nu_{\text{per}}$ for $\alpha=0$ (solid), $0.6M^{2/3}$ (dashed), $0.9M^{2/3}$ (dot dashed) with fixed $a=0.4M$.}
    \label{fig-1}
\end{figure}

The orbital frequency $\nu_\phi$, radial epicyclic frequency $\nu_r$, and vertical epicyclic frequency $\nu_\theta$, expressed in standard physical units, are given by
\begin{equation}\label{eq:38}
\nu_i = \frac{1}{2\pi} \frac{c^3}{GM} \Omega_i, \quad (i = r, \theta, \phi).
\end{equation}
Subsequently, the nodal precession frequency $\nu_\text{nod}$ and periastron precession frequency $\nu_\text{per}$  of the accretion disk can be obtained from these three fundamental frequencies by~\cite{stella1999correlations}
\begin{equation} \label{eq:39}
\nu_\text{nod} = \nu_\phi - \nu_\theta, \quad  \nu_\text{per} = \nu_\phi - \nu_r.
\end{equation}
Here, $\nu_\text{nod}$ characterizes the precession of the orbital plane, commonly referred to as the LT precession frequency, whereas $\nu_\text{per}$ describes the precession of the periastron.

In Fig.~\ref{fig-1}, we illustrates the radial profiles of the LT precession frequency $\nu_{\text{nod}}$ and the periastron precession frequency $\nu_{\text{per}}$ under different parameter conditions in order to show the influence of the black hole spin $a$ and  quantum effect $\alpha$.
Generally, the monotonic decay behaviors both for the frequencies of $\nu_{\text{nod}}$ and $\nu_{\text{per}}$ are much more significant in the inner disk, which diminish quickly along the radial distance and asymptotically vanishes at spatial infinity.
To be specific, as shown in Fig.~\ref{fig-1}(a) and Fig.~\ref{fig-1}(b), by fixing $\alpha = 0.6 M^{2/3}$, the increase of black hole spin amplifies the frequencies of $\nu_{\text{nod}}$ but suppresses the effect of $\nu_{\text{per}}$.
It should be noted that the vanishing spin $a$ causes $\nu_{\text{nod}}$ reduce to zero because $\nu_\theta=\nu_\phi$ in the Schwarzschild limit, indicating no LT precession in a non-rotating spacetime.
On the other hand, we observe the suppression effect both for $\nu_{\text{nod}}$ and $\nu_{\text{per}}$ when we increase the regular black hole parameter $\alpha$.
This suppression effect is more pronounced for periastron precession than LT precession as shown in Fig.~\ref{fig-1}(c) and Fig.~\ref{fig-1}(d).

\subsection{Constraining the parameters via MCMC simulations}
According to the RPM, the upper kHz QPOs are identical as the orbital frequency $\nu_\phi$, while the lower kHz QPOs are associated with the periastron precession frequency $\nu_{\text{per}}$. Besides, the low-frequency QPOs observed in black hole systems are also linked to the nodal precession frequency $\nu_{\text{nod}}$ \cite{Stella:1998mq}. It is noted that the RPM model assumes purely geodesic motion and neglects the effects of pressure, magnetic fields, radiation, and viscosity that could play important roles in realistic accretion disks; and this model provide no mechanism to explain how QPOs are excited and sustained to modulate the observed X-ray flux. Even so,  the RPM provides a simple but useful, geometrically motivated explanation for QPOs, capturing the key aspects of relativistic orbital dynamics. Then,
in this subsection, we use the observed QPOs data in the X-ray binary systems to Preliminarily constrain the parameters of the rotating regular black hole.
The mass of the central object and three frequencies of the selected QPO events are summarized in Tab.~\ref{tab:QPOdata}.
We shall combine these observational data with the theoretical model and employ MCMC simulations to explore the relevant physical parameter space, thereby constraining the values of the rotating regular black hole  parameters.

\begin{table}[htbp]
\centering
\caption{The mass, orbital frequencies, periastron precession frequencies, and nodal precession frequencies of QPOs from the X-ray binaries selected for analysis. The symbol $'-'$ denotes that no corresponding data was published for the QPO event.}
\label{tab:QPOdata}
\begin{tabular}{lccccc}
\hline
 & GRO J1655--40~\cite{motta2014precise} & GRS 1915+105 \cite{remillard2006x} & XTE J1550--564 \cite{remillard2002evidence}  & H1743--322 \cite{Ingram:2014ara} & XTE J1859+226 \cite{motta2022black}\\
\hline
$M$ ($M_\odot$)
    & $5.4\pm0.3$
    & $12.4^{+2.0}_{-1.8}$
    & $9.1\pm0.61$
    & $\gtrsim 9.29$
    & $7.85\pm0.46$\\
$\nu_\phi$ (Hz)
    & $441\pm2$
    & $168\pm3$
    & $276\pm3$
    & $240\pm3$
    & $227.5^{+2.1}_{-2.4}$\\
$\nu_{\text{per}}$ (Hz)
    & $298\pm4$
    & $113\pm5$
    & $184\pm5$
    & $165^{+9}_{-5}$
    & $128.6^{+1.6}_{-1.8}$\\
$\nu_{\text{nod}}$ (Hz)
    & $17.3\pm0.1$
    & --
    & --
    & $9.44\pm0.02$
    & $3.65\pm0.01$\\
\hline
\end{tabular}
\end{table}

Following the methodology outlined in Ref.~\cite{Foreman-Mackey:2012any}, we use the \texttt{emcee} algorithm to constrain the rotating regular black holes.
According to Bayes' theorem, the posterior probability of a model parameter $(\Theta)$ based on the observed data $(\mathcal{D})$ is expressed as
\begin{equation}
\mathcal{P}(\Theta \mid \mathcal{D})=\frac{\mathcal{L}(\mathcal{D} \mid \Theta) P(\Theta )}{P(\mathcal{D} )},\label{eq:20}
\end{equation}
 where $\mathcal{L}(\mathcal{D} \mid \Theta)$ represents the likelihood of the data given by the model, and $P(\Theta)$ denotes the prior distribution on the parameters, and $ P(\mathcal{D} )$ is the normalization factor. In our case, $\mathcal{D}$ represents the QPO frequencies for each X-ray binary listed in Tab.~\ref{tab:QPOdata}, while $\Theta$ represents the black hole parameters $[M,\, a/M,\, r/M,\,\alpha/M^{2/3}]$ involved in the QPOs events.
To proceed, we adopt Gaussian priors for the set of parameters $\mu_i = [M,\, a/M,\, r/M]$ as
\begin{equation}
P(\mu_i) \propto \exp \left[ -\frac{1}{2}
\left( \frac{\mu_i - \mu_{0,i}}{\sigma_i} \right)^2 \right],
\end{equation}
with parameter ranges constrained by $\mu_{0,i} < \mu_i < \mu_{\text{high},i}$, where $\sigma_i$ denotes the uncertainty associated with each parameter.
For the additional quantum effect parameter $\alpha/M^{2/3}$, we assume a uniform prior
\begin{equation}
P(\alpha/M^{2/3}) =
\begin{cases}
1, & \alpha/M^{2/3} \in [\alpha_{\text{low}},\, \alpha_{\text{high}}], \\
0, & \text{otherwise}.
\end{cases}
\end{equation}
Besides, following the orbital, periastron precession and nodal precession frequencies obtained from RPM in the above subsection, our MCMC simulations include three different parts of data from the fundamental frequencies. Thus, the likelihood function $\mathcal{L}$ is given by
\begin{eqnarray}
\log \mathcal{L} &=&\log \mathcal{L}_{\text{orb}}+\log \mathcal{L}_{\text{per}}+\log \mathcal{L}_{\text{nod}}\nonumber \\
&=&-\frac{1}{2} \sum_i
\left( \frac{\nu_{\phi,\mathrm{obs}}^i - \nu_{\phi,\mathrm{th}}^i}{\sigma_{\phi,\mathrm{obs}}^i} \right)^2
-\frac{1}{2} \sum_i
\left( \frac{\nu_{\mathrm{per,obs}}^i - \nu_{\mathrm{per,th}}^i}{\sigma_{\mathrm{per,obs}}^i} \right)^2
-\frac{1}{2} \sum_i
\left( \frac{\nu_{\mathrm{nod,obs}}^i - \nu_{\mathrm{nod,th}}^i}{\sigma_{\mathrm{nod,obs}}^i} \right)^2,
\end{eqnarray}
where $\nu_{\phi,\mathrm{obs}}^i$, $\nu_{\mathrm{per,obs}}^i$, and $\nu_{\mathrm{nod,obs}}^i$ represent the observed orbital, periastron precession, and nodal precession frequencies (see Tab.\eqref{tab:QPOdata}), while $\nu_{\phi,\mathrm{th}}^i$, $\nu_{\mathrm{per,th}}^i$, and $\nu_{\mathrm{nod,th}}^i$ correspond to their theoretical predictions from RPM, obtained by \eqref{eq:38} and \eqref{eq:39}.
$\sigma_{x,\mathrm{obs}}^i$ denote the statistical uncertainties associated with each frequency.

Using the prior distributions summarized in Tab.~\ref{table-2} and following the above setup, we generate $10^5$ random samples for each parameter to fully explore the physically admissible parameter space to determine the posterior distribution the black hole parameters $[M,\, a/M,\, r/M,\,\alpha/M^{2/3}]$.
The results of the MCMC analysis of the rotating regular black hole with the use of five observed X-ray QPO events listed in Tab.~\ref{tab:QPOdata} are shown in Fig.~\ref{fig-2} and ~\ref{fig-3}, in which the contour plots illustrate the posterior probability density distributions of the full parameter set. In each plot, the shaded regions respectively correspond to the $68\%$, $90\%$, and $95\%$ confidence levels (C.L.), giving the statistical uncertainties of the related parameters. The  dashed orange lines denote the $95\%$ confidence level for the quantum parameter $\alpha$.

The best-fit values for these parameters are also summarized in Tab.~\ref{tab:III}. These results indicates that the quantum effect parameter is less than one. Especially, its best-fit value
is from GRO J1655-40 and the upper limit is $\alpha/M^{2/3} < 0.60$ at the $95\%$ C.L.. while the upper limit from other events are close but a bit higher than 0.6. This is reasonable because from Tab.~\ref{tab:QPOdata}, GRO J1655-40 indeed have higher measurement accuracy comparing against other QPOs event. Thus, we summarize from our simulations via QPOs events that the quantum effect parameter is constrained in
\begin{equation}\label{eq:upper limit}
\alpha/M^{2/3} < 0.60
\end{equation}
at the $95\%$ C.L. It is noted that this upper limit is tighter
than $0.7014$ in our previous work \cite{Guo:2025zca}, in which we modeled the QPOs by the charged particle orbiting around the static regular black hole \eqref{metric:s} immersed in magnetic field.
Noted that QPOs were also employed to constrain the parameters or seek the preferred spacetime in more classes of regular black holes. For examples, it was investigated a class of regular black holes with a nonlinear electrodynamics (NED) theory, and the RPM of QPOs was used to constrain the NED charge parameters \cite{Banerjee:2022chn,Hazarika:2025axz}. QPO analyses have also been applied to place constraints on the quantum gravity parameter, i.e. polymeric parameter, to be less than $0.67$ by modeling a regular self-dual black holes in loop quantum gravity (LQG) ~\cite{Liu:2023vfh}. Moreover, the authors of Ref.~\cite{Jiang:2021ajk} used the observations of GRO J1655$-$40 to constrain the rotating Simpson-Visser metric, finding that the data prefer a regular black hole possessing one event horizon. In \cite{Boshkayev:2023rhr}, the authors examined the applicability of various regular black hole solutions to neutron star configurations for several QPO data sets, and found that regular black holes can describe neutron star exteriors and in most of the regular black holes metric can be better suited than the standard spherical symmetry induced by the Schwarzschild spacetime. Thus, along with all these results including ours, the X-ray observations of QPOs definitely help to testify or constrain the theory of gravity by modeling the associated regular black hole as the central object.

\begin{table}[htbp]
\centering
\caption{The Gaussian prior of the rotating regular black hole from QPOs for the X-ray Binaries.}
\label{table-2}
\begin{tabular}{lccccccccccc}
\hline
Parameters
& \multicolumn{2}{c}{GRO J1655--40}
& \multicolumn{2}{c}{GRS 1915+105}
& \multicolumn{2}{c}{XTE J1550--564}
& \multicolumn{2}{c}{H1743--322}
& \multicolumn{2}{c}{XTE J1859+226}
\\
\cline{2-11}
& $\mu$ & $\sigma$
& $\mu$ & $\sigma$
& $\mu$ & $\sigma$
& $\mu$ & $\sigma$
& $\mu$ & $\sigma$ \\
\hline
$M$ ($M_\odot$)
& 5.307 & 0.066
& 12.410 & 0.62
& 9.100  & 0.61
& 9.290  & 0.46
& 7.850  & 0.46
 \\
$a/M$
& 0.286 & 0.003
& 0.290  & 0.015
& 0.340  & 0.007
& 0.270  & 0.013
& 0.149 & 0.005
 \\
$r/M$
& 5.677 & 0.035
& 6.10  & 0.30
& 5.47  & 0.12
& 5.55  & 0.22
& 6.85  & 0.18
  \\
$\alpha /M^{2/3}$
& \multicolumn{2}{c}{Uniform [0,1]}
& \multicolumn{2}{c}{Uniform [0,1]}
& \multicolumn{2}{c}{Uniform [0,1]}
& \multicolumn{2}{c}{Uniform [0,1]}
& \multicolumn{2}{c}{Uniform [0,1]} \\
\hline
\end{tabular}
\end{table}

\begin{table}[htbp]
\centering
\caption{The best-fit values of the rotating regular black hole parameters from QPOs for the X-ray Binaries.}
\label{tab:III}
\begin{tabular}{lccccc}
\hline
Parameters
& GRO J1655--40
& GRS 1915+105
& XTE J1859+226
& H1743--322
& XTE J1550--564
 \\
\hline

$a/M$
& $0.29^{+0.00}_{-0.00}$
& $0.28^{+0.02}_{-0.02}$
& $0.15^{+0.00}_{-0.00}$
& $0.28^{+0.01}_{-0.01}$
& $0.34^{+0.01}_{-0.01}$ \\
$r/M$
& $5.68^{+0.03}_{-0.03}$
& $5.90^{+0.21}_{-0.20}$
& $6.83^{+0.11}_{-0.12}$
& $5.62^{+0.12}_{-0.14}$
& $5.46^{+0.13}_{-0.15}$ \\
$M$ ($M_\odot$)
& $5.31^{+0.06}_{-0.06}$
& $13.00^{+0.72}_{-0.73}$
& $7.91^{+0.28}_{-0.26}$
& $9.85^{+0.41}_{-0.39}$
& $8.95^{+0.39}_{-0.35}$ \\
$\alpha/ M^{2/3}$
& $< 0.60$
& $< 0.83$
& $< 0.88$
& $< 0.92$
& $< 0.86$ \\
\hline
\end{tabular}
\end{table}

\begin{figure}[h]
    \centering
    {\includegraphics[width=0.5\linewidth]{FIG2.png}}
    \caption{Corner plots showing posterior distributions for the parameters $M,\, a/M,\, r/M,\,\alpha/M^{2/3}$ of the rotating regular black hole, which is obtained by MCMC analysis for GRO J1655–40 from current observations of QPOs within the relativistic precession model.}
    \label{fig-2}
\end{figure}

\begin{figure}[h]
    \centering
    {\includegraphics[width=0.7\linewidth]{FIG3.png}}
    \caption{Corner plots showing posterior distributions for the parameters $M,\, a/M,\, r/M,\,\alpha/M^{2/3}$ of the rotating regular black hole, which are obtained by MCMC analysis  from current observations of QPOs within the relativistic precession model. Panels (a)-(d) are for QPOs events, GRS 1915+105, XTE J1859+226, H1743-322 and XTE J1550-564, respectively.}
    \label{fig-3}
\end{figure}

\section{Spin precession of test gyroscope} \label{sec:gyro}
In this section, we will examine a test gyroscope and study its spin precession frequency associated with a stationary observer in the framework of a rotating regular black hole spacetime.
For simplicity, we refer to this configuration as a ``fixed gyroscope", where the radial $r$ and polar $\theta$ coordinates remain constant with respect to an asymptotic observer at infinity.
Such a stationary observer possesses a four-velocity with the form
\begin{equation}
\begin{aligned}
u^\mu_{\text{stationary}} = u^t_{\text{stationary}} (1, 0, 0, \Omega),
\end{aligned}
\end{equation}
where $t$ denotes the time coordinate and $ \Omega= d\phi / dt$ represents the observer's angular velocity.
Consider a gyroscope moving along a timelike Killing trajectory, with its spin vector undergoing Fermi–Walker transport along a four-velocity $u=(-K^2)^{-1/2} K$, where $K$ is a timelike Killing vector field.
Since the regular black hole admits two Killing vector fields, the most general form of $K$ along which the gyroscope can move is given by
\begin{equation}
K = \partial_t + \Omega \partial_\phi.
\end{equation}
In this case, the spin precession frequency of the gyroscope is determined by the rescaled vorticity one-form of the observer congruence, which is expressed as
\begin{equation}
\begin{aligned}
\tilde{\Omega}_p = \frac{1}{2K^2} * (\tilde{K} \wedge d\tilde{K}),
\end{aligned}
\end{equation}
 where $\tilde{K}$ is the covector of $K$, $*$ denotes the Hodge dual, and $ \wedge$ is wedge product.
 The corresponding vector for the spin precession frequency~\cite{Chakraborty:2016mhx} is
\begin{equation}
\begin{aligned}\label{Eq:omega-p}
\vec{\Omega}_p =
\frac{\varepsilon_{c k l}}{2\sqrt{-g} \left(1 + 2 \Omega \frac{g_{0c}}{g_{00}} + \Omega^2 \frac{g_{cc}}{g_{00}} \right)}
\left[
\left( \frac{g_{0c,k}}{g_{00}} - \frac{g_{0c} g_{00,k}}{g_{00}^2} \right)
+ \Omega \left( \frac{g_{cc,k}}{g_{00}} - \frac{g_{cc} g_{00,k}}{g_{00}^2} \right)
+ \Omega^2 \left( \frac{g_{0c}}{g_{00}} \frac{g_{cc,k}}{g_{00}} - \frac{g_{cc} g_{0c,k}}{g_{00}^2} \right)
\right] \partial_l,
\end{aligned}
\end{equation}
where $\varepsilon_{c k l}(c,k,l=1,2,3)$ is the Levi-Civita symbol.

Substituting the regular black hole metric~\eqref{eq:matric_r} into the general spin precession expression~\eqref{Eq:omega-p}, the expression can be simplified to
\begin{equation}\label{eq:14}
\vec{\Omega}_p =
\frac{(C_1 \hat{r} + C_2 \hat{\theta})}
{2 \sqrt{-g} \left( 1 + 2 \Omega \frac{g_{t\phi}}{g_{tt}} + \Omega^2 \frac{g_{\phi\phi}}{g_{tt}} \right)},
\end{equation}
with
\begin{equation}
\begin{aligned}\label{eq:15}
C_1 &= -\sqrt{g_{rr}} \left[
\left( \frac{g_{t\phi,\theta}}{g_{tt}} - \frac{g_{t\phi}}{g_{tt}^2} g_{tt,\theta} \right)
+ \Omega \left( \frac{g_{\phi\phi,\theta}}{g_{tt}} - \frac{g_{\phi\phi}}{g_{tt}^2} g_{tt,\theta} \right) \right. \left.
+ \Omega^2 \left( \frac{g_{t\phi}}{g_{tt}} \frac{g_{\phi\phi,\theta}}{g_{tt}} - \frac{g_{\phi\phi}}{g_{tt}} \frac{g_{t\phi,\theta}}{g_{tt}} \right)
\right],
\\
C_2 &= \sqrt{g_{\theta\theta}} \left[
\left( \frac{g_{t\phi,r}}{g_{tt}} - \frac{g_{t\phi}}{g_{tt}^2} g_{tt,r} \right)
+ \Omega \left( \frac{g_{\phi\phi,r}}{g_{tt}} - \frac{g_{\phi\phi}}{g_{tt}^2} g_{tt,r} \right) \right. \left.
+ \Omega^2 \left( \frac{g_{t\phi}}{g_{tt}} \frac{g_{\phi\phi,r}}{g_{tt}} - \frac{g_{\phi\phi}}{g_{tt}} \frac{g_{t\phi,r}}{g_{tt}} \right)
\right].
\end{aligned}
\end{equation}
Similar to the case of~\cite{Wu:2023wld}, the observation is valid only in a restricted range of the angular velocity $\Omega(r,\theta)$
\begin{equation}
\begin{aligned}\label{eq:16}
 \Omega_-<\Omega(r, \theta) <\Omega_+,
\end{aligned}
\end{equation}
with the bound defined by
\begin{equation}\label{eq:34}
\Omega_{\pm} = \frac{-g_{t\phi} \pm \sqrt{g_{t\phi}^2 - g_{\phi\phi} g_{tt}}}{g_{\phi\phi}}.
\end{equation}
Due to the non-zero angular velocity of the gyroscope, its general relativistic spin precession arises from two distinct sources: the LT precession, induced by the rotation of the black hole, and the geodetic precession, caused by the curvature of the surrounding spacetime.
In the following discussion, we will analyze the LT precession, the geodesic precession, and the general spin precession within the context of regular black hole spacetimes. In this section, we shall work in the unit with $G=c=1$, and set $M=1$.

\subsection{Lense-Thirring precession frequency}

When the angular velocity vanishes, i.e. $\Omega = 0$, the gyroscope is held by a static observer, who remains at rest with respect to the coordinate system.
Such a static observer can exist only outside the ergoregion, since remaining static within it would require infinite acceleration and is therefore unphysical.
In this case, the four-velocity of the static observer is  $u^\mu_{\text{stationary}} = (u^t_{\text{stationary}}, 0, 0, 0)$ and the Killing vector reads $K = \partial_t$.
Consequently, the rescaled spin precession frequency $\vec{\Omega}_p$ reduces to the LT precession frequency $\vec{\Omega}_{\text{LT}}$~\cite{Chakraborty:2013naa}.
Its vectorial expression is given by
\begin{equation}
\vec{\Omega}_{\text{LT}} = \frac{1}{2\sqrt{-g}}
\left\{
-\sqrt{g_{rr}} \left( g_{t\phi,\theta} - \frac{g_{t\phi}}{g_{tt}} g_{tt,\theta} \right) \hat{r}
+ \sqrt{g_{\theta\theta}} \left( g_{t\phi,r} - \frac{g_{t\phi}}{g_{tt}} g_{tt,r} \right) \hat{\theta}
\right\}.
\end{equation}
The magnitude of the LT precession frequency can be obtained as
\begin{equation}\label{eq:LT}
\Omega_{\text{LT}} = \frac{1}{2 \sqrt{-g}}
\sqrt{
g_{rr} \left( g_{t\phi,\theta} - \frac{g_{t\phi}}{g_{tt}} g_{tt,\theta} \right)^2
+ g_{\theta\theta} \left( g_{t\phi,r} - \frac{g_{t\phi}}{g_{tt}} g_{tt,r} \right)^2
}.
\end{equation}

Fig.~\ref{fig-4} illustrates the radial profiles of the LT precession frequency $(\Omega_{\mathrm{LT}})$, which increases as the stationary gyroscope approaches the rotating regular black hole and diverges at the ergoregion.
For the case with a fixed parameter $\alpha = 0.6\,M^{2/3}$ and an inclination angle $\theta = \pi/2$, $\Omega_{\mathrm{LT}}$ increases significantly with the spin parameter $a$, indicating that higher spin strengthens the frame-dragging effect and lead to stronger distortions of local inertial frames near the ergosphere.
When we fix $a = 0.2M$ and $\alpha = 0.6\,M^{2/3}$, varying the polar angle $\theta$ from the rotation axis ($\theta = 0$) toward the equatorial plane ($\theta = \pi/2$) shifts the curves to the right, revealing the pronounced angular dependence of the frame-dragging effect: observers near the equatorial plane experience stronger frame-dragging than those closer to the rotation axis.
These observations are reasonable, since the LT precession frequency primarily reflects the rotation of the black hole, and both the black hole spin and the gyroscope inclination are tightly correlated with the spin itself.
From this perspective, an increase in the regular black hole parameter does not necessarily enhance the LT frequency.
Indeed, enlarging the parameter $\alpha$ markedly suppresses the LT precession frequency and shifts the corresponding curves toward smaller radial coordinates.
This behavior reflects that the regular black hole parameter weakens the frame-dragging effect.

\begin{figure}[h]
    \centering
    {\includegraphics[width=0.9\linewidth]{FIG4.png}}
    \caption{Lense-Thirling precession frequency as a function of radial coordinate for rotating regular black hole. In plot (a), we check the effect of the black hole spin with $\alpha=0.6 M^{2/3},\ \theta=\pi/2$; In plot (b), the figure shows the effect of the angle with $a=0.2M, \alpha=0.6 M^{2/3}$; And in plot (c), we check the effect of the regular black hole parameter $\alpha$ with fixed $a=0.2M,\ \theta=\pi/2$.}
    \label{fig-4}
\end{figure}

\subsection{Geodetic precession frequency}

When $a=0$, the black hole loses its rotational angular momentum, and the rotating regular black hole described by metric \eqref{eq:matric_r} reduces to the static configuration~\eqref{metric:s}. In this limit, the LT precession frequency \eqref{eq:LT} vanishes, whereas the spin precession frequency \eqref{Eq:omega-p} remains finite. The surviving nonzero contribution to the spin precession, arising from the curvature of spacetime, is known as geodetic precession, and is given by
\begin{equation}
\vec{\Omega}_p \big|_{a=0} =
\frac{1}{2 \sqrt{-g} \left(1 + \Omega^2 \frac{g_{\phi\phi}}{g_{tt}} \right)}
\left[
- \Omega \sqrt{g_{rr}} \left( g_{\phi\phi,\theta} - \frac{g_{\phi\phi}}{g_{tt}} g_{tt,\theta} \right) \hat{r}
+ \Omega \sqrt{g_{\theta\theta}} \left( g_{\phi\phi,r} - \frac{g_{\phi\phi}}{g_{tt}} g_{tt,r} \right) \hat{\theta}
\right].
\end{equation}
 In this scenario, the spherical symmetry of the black hole permits the choice $\theta=\pi/2$, selecting observers restricted to the equatorial plane. Under this condition, the geodetic precession frequency reduces to
\begin{equation}
\Omega_p \big|_{a=0,\theta=\pi/2} =
\frac{ \Omega \sqrt{g_{\theta\theta}} \left( g_{\phi\phi,r} - \frac{g_{\phi\phi}}{g_{tt}} g_{tt,r} \right) }
{ 2 \sqrt{-g} \left(1 + \Omega^2 \frac{g_{\phi\phi}}{g_{tt}} \right) }=\frac{\Omega  \left(3 \alpha ^3-3 r^3+r^4 e^{\frac{\alpha
   ^3}{r^3}}\right)}{r^3 \left(r e^{\frac{\alpha ^3}{r^3}} \left(r^2 \Omega
   ^2-1\right)+2\right)},
\end{equation}
indicating that a gyroscope moving in a static regular black hole spacetime still experiences precession. For a gyroscope following a geodesic trajectory in the equatorial plane, the angular frequency coincides with the orbital angular velocity~\cite{Glendenning:1993di}
\begin{equation}
\Omega  = \left. \frac{d\phi}{dt} \right|_{a=0,\, \theta = \pi/2}=\frac{\sqrt{e^{-\frac{\alpha ^3}{r^3}} \left(r^3-3 \alpha ^3\right)}}{r^3}.
\end{equation}
Substituting this into the expression for geodetic precession, we have
\begin{equation}
\Omega_p \big|_{a=0,\Omega= \left. \frac{d\phi}{dt} \right|_{a=0,\, \theta = \pi/2}}=\frac{\sqrt{e^{-\frac{\alpha ^3}{r^3}} \left(r^3-3 \alpha ^3\right)}}{r^3}.
\end{equation}
The expression above provides the precession frequency in the Copernican coordinates, measured with respect to the proper time $\tau$. Considering the relation between coordinate time $t$ and proper time $\tau$
\begin{equation}
d\tau = \sqrt{1-\frac{2 e^{-\frac{\alpha ^3}{r^3}}}{r}} dt,
\end{equation}
the geodetic precession frequency in coordinate time takes the form
\begin{equation}\label{eq:41}
\Omega_{\text{geodetic}} =\frac{\sqrt{1-\frac{2 e^{-\frac{\alpha ^3}{r^3}}}{r}} \sqrt{e^{-\frac{\alpha
   ^3}{r^3}} \left(r^3-3 \alpha ^3\right)}}{r^3}.
\end{equation}

As mentioned earlier, the geodetic precession frequency arises from the curvature of spacetime.
Consequently, when the regular black hole parameter vanishes, Eq.~\eqref{eq:41} reduces to the geodetic precession frequency of a Schwarzschild black hole.
As shown in \red{ Fig.~\ref{fig-5}}, increasing this parameter $\alpha$ suppresses the geodetic precession frequency of the regular black hole relative to the Schwarzschild result, leading to larger deviations.
This effect is even more clearly illustrated in the right panel, and we observe that the deviation becomes increasingly pronounced near the black hole horizon.

\begin{figure}[h]
    \centering
{\includegraphics[width=0.8\linewidth]{FIG5.png}}
    \caption{The geodetic precession as a function of radial coordinate for a static regular black hole. The left panel desplays $\Omega_{\text{geodetic}}$ for different values of $\alpha$, while the right panel represents the deviation of $\Omega_{\text{geodetic}}$ from the corresponding Schwarzschild black hole with the same parameters as those in the left panel.}
    \label{fig-5}
\end{figure}

\subsection{General spin precession frequency}
Now we move on to investigate the general spin precession frequency in the spacetime of rotating regular black hole. For a gyroscope attached to a static observer, its angular velocity is constrained by Eq.~\eqref{eq:16}. Therefore, by introducing a parameter $0 < k < 1$, the angular velocity $\Omega$ can be expressed as
\begin{equation}\label{eq:26}
\Omega = k\,\Omega_+ + (1-k)\,\Omega_- = \frac{(2k-1)\,\sqrt{g_{t\phi}^2 - g_{tt}g_{\phi\phi}} - g_{t\phi}}{g_{\phi\phi}}
\end{equation}
where $\Omega_{\pm}$ are defined in Eq.~\ref{eq:34}.

By substituting Eq.~\eqref{eq:26} into Eq.~\eqref{eq:14}, we derive a general form for the spin precession frequency
\begin{equation}
\Omega_p =
\frac{ \sqrt{ C_1^2 + C_2^2\, g_{tt} g_{\phi\phi} } }
{ 8 \sqrt{-g}\, (k - 1)\, k \left( g_{t\phi}^2 - g_{tt} g_{\phi\phi} \right) }
\end{equation}
where $C_1$ and $C_2$ are specified in Eq.~\eqref{eq:15}.

Fig.~\ref{fig-6} shows the magnitude of the spin precession frequency in the rotating regular black hole spacetime for representative values of $k$, specifically $k = 0.1$, $0.5$, and $0.9$.
For $k=0.1$ and $k=0.9$ (Fig.~\ref{fig-6}(a) and Fig.~\ref{fig-6}(c)), we observe that the spin precession frequency $\Omega_p$ diverges near the black hole horizon for all inclination angles.
It is noteworthy that when $k=0.5$ (Fig.~\ref{fig-6}(b)), the observer corresponds to a zero angular momentum observer (ZAMO).
In this case, the spin precession frequency remains finite from any direction, because these gyroscopes are effectively non-rotating relative to the local spacetime geometry and the stationary observers.

\begin{figure}[h]
    \centering
 {\includegraphics[width=0.9\linewidth]{FIG6.png}}
    \caption{The effect of $\theta$ on the spin precession frequency $\Omega_p$ for different $k$ with fixed $a=0.2M$ and $\alpha=0.6M^{2/3}$ in rotating regular black hole spacetime.  Panels (a)-(c) correspond to  $k=0.1,~0.5,$ and $0.9$, respectively.}
    \label{fig-6}
\end{figure}

\section{Conclusion}\label{sec:conclusion}

In this work, we investigated the spin precession effects and quasi-periodic oscillations (QPOs) in the spacetime of a rotating regular black hole and used X-ray binary observations to constrain the deviation parameter of this regular solution. For test-particle orbits, both the periastron precession frequencies and the nodal precession frequencies were found to be suppressed with increasing $\alpha$, and their radial profiles exhibited  deviations from the Kerr case, particularly in the regions close to the black hole.

By performing MCMC fitting of the QPO data from five X-ray binary systems (GRO\,J1655–40, GRS\,1915+105, XTE\,J1859+226, H1743–322, and XTE\,J1550–564), we obtained stringent constraints on the parameters of the rotating regular black hole model. For all sources, the deviation parameter satisfied $\alpha/M^{2/3}<1$, with the tightest constraint coming from GRO\,J1655–40, where we found $\alpha/M^{2/3}<0.60$ at the $95\%$ confidence level. Moreover, the best-fit values of the black hole mass $M$, spin parameter $a$, and orbital radius $r$ for these systems showed only minimal deviations from those predicted by the classical Kerr spacetime. This indicated that the QPO observations were in strong agreement with the Kerr metric, thereby the data are consistent with the Kerr metric within the uncertainties, and any deviation parameterized by $\alpha/M^{2/3}$ must be smaller than the obtained upper limit. Indeed, across all five X-ray binaries, the deviation parameter $\alpha$ was constrained to within  $1$, and no significant non-Kerr behavior was detected. This implied that any possible quantum-gravity-induced deviations in these systems must have been extremely small. Consequently, our observations provided a robust upper bound on the deviation parameter $\alpha$ in the rotating regular black hole model, substantially narrowing the allowed parameter space for such quantum gravity corrections.

Furthermore, under the constraint $\alpha/M^{2/3}<0.60$, we analyzed the spin precession of a test gyroscope in the regular black hole background and derived analytical expressions for the Lense–Thirring precession frequency, geodetic precession frequency, and general spin precession frequency. We found that a nonzero $\alpha$ systematically suppressed the magnitudes of these precession frequencies. In addition, this influence was found to be anisotropic: observers at different inclination angles relative to the equatorial plane experienced distinct modifications in precession behavior.

Looking toward the future, this methodology can be further extended and refined to test the nature of regular black holes more precisely. First, applying this approach to a larger sample of X-ray binaries and to black hole systems across different mass scales will help verify the robustness and universality of the constraints on $\alpha$, especially for the time-evolving accretion structure \cite{Zhao:2025,Guo:2025yin} . Second, exploring the QPO signatures predicted by other types of regular black hole solutions beyond the specific metric examined here may reveal distinctive observational differences, providing a potential way to discriminate between alternative models. Moreover, next-generation X-ray observatories, such as eXTP \cite{eXTP:2018anb} and Athena \cite{Nandra:2013jka}, with their significantly improved timing precision and sensitivity, will deliver higher-quality QPO data, thereby enabling tighter bounds on the regular parameter $\alpha$ and other possible deviations from Kerr. Finally, it will be valuable to generalize the QPO frequency and spin precession models to more complex and extreme scenarios, such as non-axisymmetric or tilted accretion disk configurations and nearly extremal spins, to assess the limits of the proposed framework and ensure its applicability in realistic astrophysical environments.

\section*{Acknowledgments}
This work is partly supported by Natural Science Foundation of China under Grants No.12375054 and 12405067.
Meng-He Wu is also sponsored by Natural Science Foundation of Sichuan (No. 2025ZNSFSC0876).
H.G. is supported by the Institute for Basic Science (Grant No. IBS-R018-Y1).

{\bf Conflict of Interest}  The authors declare that they have no conflict of interest.

\bibliographystyle{JHEP}
\bibliography{refers}

\end{document}